\documentclass[conference,a4paper]{APSIPA2025}
\usepackage{amsmath}
\usepackage{graphicx}
\usepackage{multirow}
\usepackage{threeparttable}
\usepackage[backend=biber,bibstyle=ieee,citestyle=numeric-comp]{biblatex}

\usepackage{booktabs}
\usepackage{multirow}
\usepackage{caption}
\usepackage{tabularx}
\usepackage{titlesec}
\usepackage{lmodern}
\usepackage{lipsum}
\usepackage{svg}
\usepackage{graphicx}
\usepackage[colorlinks=true, linkcolor=black, citecolor=black, urlcolor=black]{hyperref}

\titleformat{\section}
  [block]
  {\centering\bfseries\large}
  {\thesection.}{0.5em}{}

\renewcommand\thesubsection{\Alph{subsection}}
\titleformat{\subsection}
  [block]
  {\itshape\bfseries}
  {\thesubsection.}{0.25em}{}

\usepackage{geometry}
\usepackage{fancyhdr}

\fancypagestyle{firststyle}{
  \fancyhf{}                                      
  \fancyhead[C]{2025 Asia Pacific Signal and Information Processing\
   Association Annual Summit and Conference (APSIPA ASC)}
}

\fancypagestyle{emptyheader}{
  \fancyhf{}                  
  \fancyfoot[C]{\thepage}     
}

\begin{document}

\title{Analysis of Speaker Verification Performance Trade-offs with Neural Audio Codec Transmission}

\author{
\authorblockN{
Nirmalya Mallick Thakur\authorrefmark{1} and
Jia Qi Yip\authorrefmark{2}\authorrefmark{3} and
Eng Siong Chng\authorrefmark{2} 
}

\authorblockA{
\authorrefmark{1}
Indian Institute of Science Education and Research, Bhopal \\
}

\authorblockA{
\authorrefmark{2}
Nanyang Technological University, Singapore \\
}

\authorblockA{
\authorrefmark{3}
Menlo Research \\
\href{mailto:nirmalya23@iiserb.ac.in}{nirmalya23@iiserb.ac.in}
}
}

\maketitle
\thispagestyle{firststyle}    
\pagestyle{emptyheader}       

\begin{abstract}

Neural audio codecs (NACs) have made significant advancements in recent years and are rapidly being adopted in many audio processing pipelines. However, they can introduce audio distortions which degrade speaker verification (SV) performance. This study investigates the impact of both traditional and neural audio codecs at varying bitrates on three state-of-the-art SV models evaluated on the VoxCeleb1 dataset. Our findings reveal a consistent degradation in SV performance across all models and codecs as bitrates decrease. Notably, NACs do not fundamentally break SV performance when compared to traditional codecs. They outperform Opus by 6-8\% at low-bitrates (\textless\ 12 kbps) and remain marginally behind at higher bitrates ($\approx$ 24 kbps), with an EER increase of only 0.4-0.7\%. The disparity at higher bitrates is likely due to the primary optimization of NACs for perceptual quality, which can inadvertently discard critical speaker-discriminative features, unlike Opus which was designed to preserve vocal characteristics. Our investigation suggests that NACs are a feasible alternative to traditional codecs, especially under bandwidth limitations. To bridge the gap at higher bitrates, future work should focus on developing speaker-aware NACs or retraining and adapting SV models.
\end{abstract}


\section{Introduction}



Speaker verification (SV) is the task of determining whether a given audio sample belongs to a claimed speaker identity. In many cases, audio has to be transmitted to a server where SV is performed. Efficient audio compression is essential for numerous applications, particularly transmission in bandwidth-limited environments~\cite{10445746} and real-time communication systems~\cite{kondoz2005digital}. These constraints create a trade-off between transmission bandwidth and performance, where in general, a better audio codec is one that gives higher performance at a given bandwidth.


Neural audio codecs (NACs) have recently gained much attention for their excellent performance and rapid development in recent years~\cite{soundstream,encodec,dac}. They are autoencoder networks with a quantizer at their center. Much of the rapid advancement has come from the residual vector quantization (RVQ) method~\cite{soundstream}. Quantizers are often divided into multiple residual layers to balance between bandwidth efficiency and compression quality. Nevertheless, it has been shown that NACs do not always outperform traditional methods on all downstream tasks~\cite{dsab}. In a given system consisting of a transmission step and a SV step, neural audio codecs also need to be evaluated against a traditional codec for bandwidth and performance trade-offs.


The interaction between audio compression and SV presents unique challenges. Non-neural codecs like Opus are hand-engineered for efficient transmission, while NACs are trained to optimize perceptual quality. This fundamental difference in their design philosophy motivates an evaluation of both methods for SV related tasks. The performance trends across varying bitrates as shown in Figure 1 highlights the strengths and drawbacks of both approaches.

\begin{figure}[t]
  \centering
  \includesvg[width=\linewidth]{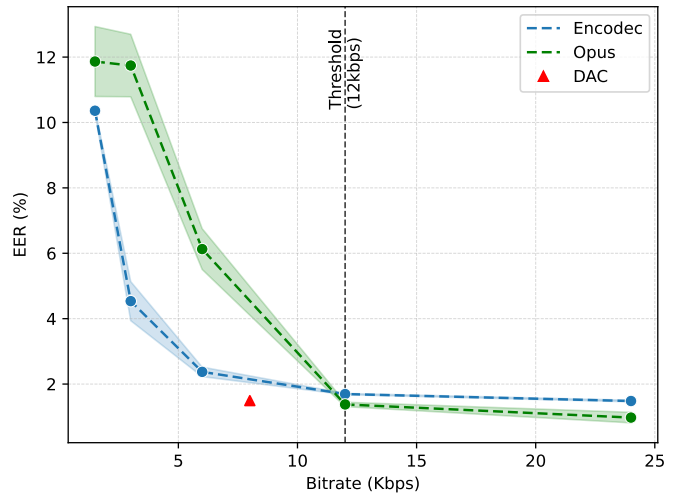}
  \caption{EER versus bitrate for NACs and Opus on the VoxCeleb1-O (cleaned) split. The shaded regions indicate standard deviation across the three SV models.}
  \label{fig:nac_versus_opus_comparison}
  \vspace{-5pt}
\end{figure}

In this work, we examine the performance trade-offs between NACs and traditional codecs on SV tasks. We make the following contributions:

\begin{itemize}
\item Evaluate the impact of audio compression on SV performance by testing three state-of-the-art SV models (ECAPA-TDNN~\cite{ecapa-tdnn}, CAM++~\cite{cam++}, and ERes2Net-Large~\cite{eres2net}) on the VoxCeleb1~\cite{voxceleb} dataset across different bandwidths.
\item Present the performance difference between neural (Encodec~\cite{encodec}, DAC~\cite{dac}) and traditional codecs (Opus~\cite{opus}) on downstream SV tasks by testing them across a wide range of compression levels.
\item Demonstrate that SV performance degrades predictably with bitrate by quantifying our results through metrics.
\item Through visualization of speaker embeddings, we present insights into how audio compression affects the speaker embedding space.
\end{itemize}

Overall, our findings suggest that NACs do not fundamentally break SV systems. NACs outperform Opus in bandwidth-constrained scenarios, while Opus shows only a marginal advantage at higher bitrates. Therefore, NACs represent a robust choice for SV applications, although a careful evaluation of the compression-performance trade-off needs to be made.






\section{Methodology}
In this study, we investigate the impact of audio compression on the performance of SV models by processing the audio samples through different codecs at varying bitrates. In the evaluation pipeline, we first process the baseline VoxCeleb1 dataset using codecs at different bitrates. The performance of the SV models is then evaluated on the degraded samples using standard recipes from 3D-Speaker~\cite{chen20243d} as shown in Figure 2.

\subsection{Speaker Verification Models}
For this study, we used three state-of-the-art SV models with different numbers of parameters and memory requirements:

\begin{itemize}
  \item ECAPA-TDNN: A time-delay neural networks (TDNN) based architecture with 1D Res2Net modules, SE blocks, and Multi-scale Feature Aggregation.
  \item CAM++: Based on D-TDNN with context-aware masking and multi-granularity pooling. It achieves ECAPA-level performance with lower computational cost and faster inference.
  \item ERes2Net-Large: Enhances Res2Net with local and global feature fusion using attention-based modules.
\end{itemize}

\begin{table}[h]
\centering
\caption{Different speaker verification models and their parameter size}
\begin{tabular}{lcc}
\toprule
\textbf{Model} & \textbf{Architecture} & \textbf{Params(M)} \\
\midrule
ECAPA-TDNN & TDNN + SE-Res2Blocks & 20.8M \\
CAM++ & D-TDNN + Context Masking & 7.2M \\
ERes2Net-Large & E-Res2Net + LFF + GFF & 22.46M \\
\bottomrule
\end{tabular}
\label{tab:model_params}
\end{table}

All the models were evaluated using pre-trained checkpoints from ModelScope~\cite{modelscope}, with no fine-tuning performed for this study.

\subsection{Codec based Compression Framework}

We assessed the effect of audio compression through both traditional and neural audio codecs (NACs):
\begin{itemize}
    \item Neural Audio Codecs (NACs): Encodec (1.5, 3, 6, 12, 24 kbps) and Descript Audio Codec or DAC (8 kbps) were evaluated separately at supported bitrate settings.
    \item Traditional Codecs: Opus was applied via \textit{ffmpeg} at comparable bitrates (1.5, 3, 6, 12, 24 kbps) to serve as a non-neural baseline.
\end{itemize}

Each codec was used to compress and decompress the audio, simulating an end-to-end transmission pipeline. This allowed us to analyze codec-specific artifacts under varying bitrate constraints.

\begin{figure}[h]
  \centering
  \includegraphics[width=\linewidth]{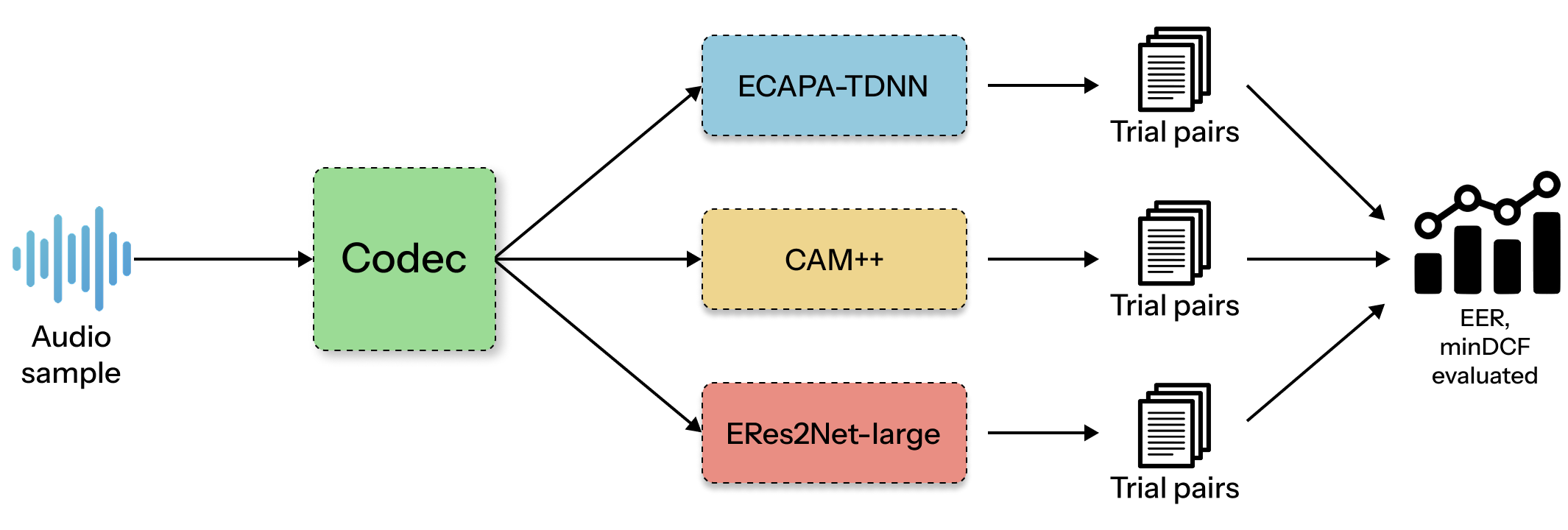}
  \caption{Experimental pipeline overview}
  \label{fig:pipeline_image}
\end{figure}

\subsection{Dataset and Processing Pipeline}
We used the test split of the VoxCeleb1 dataset in our analysis. As shown in Table II, there are three tasks on VoxCeleb1, and the last two tasks have more trials. To simulate different levels of compression, the original audio files were processed through each codec at multiple bitrate settings and stored in separate directories. All the files were resampled to 16 kHz to match the default sampling rate of the SV models. Finally, compressed samples were scored using pre-configured 3D-Speaker recipes. The trials were registered as enrollment–test pairs and verification was performed using cosine similarity. 

We report the final results using two performance metrics — the equal error rate (EER) and the minimum of the normalized detection cost function (MinDCF) with the settings of $P_{target} = 0.01$
and $C_{fa} = C_{miss} = 1$.

\begin{table}[h]
\centering
\caption{Tasks on VoxCeleb1 dataset. Here 'O' denotes 'original', 'E' denotes 'extended', and 'H' denotes 'hard'}
\begin{tabular}{lccc}
\toprule
& VoxCeleb1-O & VoxCeleb1-E & VoxCeleb1-H \\
\midrule
Speakers & 40 & 1251 & 1251 \\
Trials & 37,611 & 579,818 & 550,894 \\
\bottomrule
\end{tabular}
\label{tab:voxceleb_dataset}
\end{table}

\begin{table*}
    \centering
    \caption{Performance comparison of all three speaker verification models on the VoxCeleb1-O (cleaned) test set after being processed through different Neural Audio Codecs (NACs) at varying bitrates.}
    \label{tab:performance_comparison_vox1_O}
    \begin{tabularx}{0.85\textwidth}{>{\centering\arraybackslash}X >{\centering\arraybackslash}X 
                                     >{\centering\arraybackslash}X >{\centering\arraybackslash}X 
                                     >{\centering\arraybackslash}X}
        \toprule
        \multirow{2}{*}{\textbf{Codec}} & \multirow{2}{*}{\textbf{Bitrate (kbps)}} & \textbf{ECAPA-TDNN} & \textbf{CAM++} & \textbf{ERes2Net-Large} \\
        & & \textbf{EER(\%)/MinDCF} & \textbf{EER(\%)/MinDCF} & \textbf{EER(\%)/MinDCF} \\
        \midrule
        Baseline & --- & 0.86/0.0993 & 0.73/0.0910 & 0.57/0.0567 \\
        \midrule
        \multirow{5}{*}{Encodec} & 1.5 & 10.29/0.8017 & 10.39/0.8214 & 10.39/0.8214 \\
        & 3 & 4.12/0.4458 & 4.27/0.4323 & 5.22/0.5142 \\
        & 6 & 2.28/0.2848 & 2.30/0.2613 & 2.54/0.3001 \\
        & 12 & 1.72/0.2258 & 1.68/0.1986 & 1.69/0.2150 \\
        & 24 & 1.51/0.1940 & 1.46/0.1643 & 1.47/0.1765 \\
        \midrule
        DAC & 8 & 1.39/0.1893 & 1.49/0.1945 & 1.39/0.1894 \\
        \midrule
        \multirow{5}{*}{Opus} & 1.5 & 11.75/0.8325 & 10.85/0.8227 & 12.99/0.8725 \\
        & 3 & 11.61/0.8294 & 10.86/0.8298 & 12.76/0.8561 \\
        & 6 & 6.38/0.6034 & 5.42/0.5796 & 6.59/0.5836 \\
        & 12 & 1.44/0.1881 & 1.39/0.2103 & 1.30/0.1991\\
        & 24 & 1.10/0.1258 & 1.03/0.1394 & 0.79/0.0985 \\
        \bottomrule
    \end{tabularx}
\end{table*}

\begin{table*}
    \centering
    \caption{Performance comparison of all three speaker verification models on the VoxCeleb1-E (cleaned) test set after being processed through different Neural Audio Codecs (NACs) at varying bitrates.}
    \label{tab:performance_comparison_vox1_E}
    \begin{tabularx}{0.85\textwidth}{>{\centering\arraybackslash}X >{\centering\arraybackslash}X 
                                     >{\centering\arraybackslash}X >{\centering\arraybackslash}X 
                                     >{\centering\arraybackslash}X}
        \toprule
        \multirow{2}{*}{\textbf{Codec}} & \multirow{2}{*}{\textbf{Bitrate (kbps)}} & \textbf{ECAPA-TDNN} & \textbf{CAM++} & \textbf{ERes2Net-Large} \\
        & & \textbf{EER(\%)/MinDCF} & \textbf{EER(\%)/MinDCF} & \textbf{EER(\%)/MinDCF} \\
        \midrule
        Baseline & --- & 0.96/0.1112 & 0.89/0.0996 & 0.79/0.0848 \\
        \midrule
        \multirow{5}{*}{Encodec} & 1.5 & 11.20/0.8463 & 11.67/0.8648 & 11.67/0.8650 \\
        & 3 & 4.67/0.4754 & 4.95/0.4845 & 6.24/0.5723 \\
        & 6 & 2.60/0.2869 & 2.75/0.2996 & 3.25/0.3373 \\
        & 12 & 1.84/0.2041 & 1.87/0.2103 & 2.03/0.2179 \\
        & 24 & 1.64/0.1874 & 1.65/0.1869 & 1.73/0.1880 \\
        \midrule
        DAC & 8 & 1.64/0.1865 & 1.58/0.1894 & 1.64/0.1865 \\
        \midrule
        \multirow{5}{*}{Opus} & 1.5 & 11.72/0.8457 & 10.70/0.8386 & 13.15/0.9026 \\
        & 3 & 11.56/0.8442 & 10.72/0.8376 & 12.92/0.8858 \\
        & 6 & 6.63/0.6050 & 5.76/0.5616 & 7.05/0.6432 \\
        & 12 & 1.56/0.1720 & 1.52/0.1716 & 1.53/0.1684 \\
        & 24 & 1.10/0.1223 & 1.06/0.1148 & 0.97/0.1031 \\
        \bottomrule
    \end{tabularx}
\end{table*}

\begin{table*}
    \centering
    \caption{Performance comparison of all three speaker verification models on the VoxCeleb1-H (cleaned) test set after being processed through different Neural Audio Codecs (NACs) at varying bitrates.}
    \label{tab:performance_comparison_vox1_H}
    \begin{tabularx}{0.85\textwidth}{>{\centering\arraybackslash}X >{\centering\arraybackslash}X 
                                     >{\centering\arraybackslash}X >{\centering\arraybackslash}X 
                                     >{\centering\arraybackslash}X}
        \toprule
        \multirow{2}{*}{\textbf{Codec}} & \multirow{2}{*}{\textbf{Bitrate (kbps)}} & \textbf{ECAPA-TDNN} & \textbf{CAM++} & \textbf{ERes2Net-Large} \\
        & & \textbf{EER(\%)/MinDCF} & \textbf{EER(\%)/MinDCF} & \textbf{EER(\%)/MinDCF} \\
        \midrule
        Baseline & --- & 1.85/0.1759 & 1.76/0.1727 & 1.51/0.1474 \\
        \midrule
        \multirow{5}{*}{Encodec} & 1.5 & 19.46/0.9251 & 20.19/0.9286 & 20.19/0.9286 \\
        & 3 & 8.91/0.6408 & 9.32/0.6524 & 11.17/0.7223 \\
        & 6 & 5.12/0.4163 & 5.33/0.4261 & 6.07/0.4633 \\
        & 12 & 3.66/0.3228 & 3.74/0.3265 & 3.93/0.3330 \\
        & 24 & 3.27/0.2937 & 3.30/0.2990 & 3.37/0.2960 \\
        \midrule
        DAC & 8 & 3.26/0.2885 & 3.25/0.2950 & 3.26/0.2885 \\
        \midrule
        \multirow{5}{*}{Opus} & 1.5 & 19.30/0.9210 & 18.43/0.9325 & 21.34/0.9646\\
        & 3 & 19.14/0.9209 & 18.40/0.9314 & 20.98/0.9459 \\
        & 6 & 12.04/0.7452 & 10.74/0.7254 & 12.58/0.7747\\
        & 12 & 3.08/0.2751 & 3.01/0.2875 & 2.98/0.2620 \\
        & 24 & 2.14/0.1972 & 2.06/0.2006 & 1.82/0.1703 \\
        \bottomrule
    \end{tabularx}
\end{table*}

\section{Results}
We report the final performance of all speaker verification (SV) models across different audio codecs and bitrates for the three different splits of VoxCeleb1. Tables III, IV, and V present the EER and MinDCF values for VoxCeleb1-O, VoxCeleb1-E, and VoxCeleb1-H test splits, respectively. Each table organizes the results by codec and bitrate, allowing comparisons to be made across different SV models at the same bitrate. 

\subsection{Overall Performance Trends}
All three test splits show an increasing EER trend with decreasing bitrates across codecs and SV models. As shown in Figure 3 for Encodec on the VoxCeleb1-O split, this pattern also holds for other codecs and splits. Figure 1 reveals that while Opus shows a slight advantage beyond a certain threshold (> 12 kbps), the performance gains are quite modest (around 0.4-0.7\%). At lower bitrates, NACs significantly outperform traditional codecs, achieving 6-8\% better EER at 3 kbps, with the performance gap being most notable in the 3-12 kbps range. At 24 kbps, while Opus achieves near-baseline results, Encodec remains competitive, with only marginally higher degradation. DAC at 8 kbps achieves similar or even slightly better results compared to Encodec at 24 kbps. These findings demonstrate that NACs do not compromise SV tasks, maintaining competitive performance at higher bitrates while providing significantly better results under bandwidth-constrained scenarios.

\begin{figure}[h]
  \centering
  \includesvg[width=\linewidth]{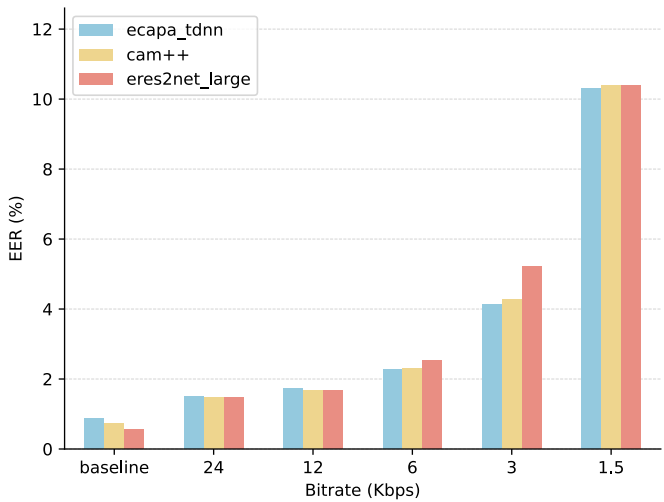}
  \caption{EER comparison across varying bitrates for Encodec evaluated using all three SV models on the VoxCeleb1-O (cleaned) split.}
  \label{fig:eer_comparision}
\end{figure}

\subsection{Embedding Space Analysis}
    We extracted speaker embeddings from 50 different speakers with five utterances each, using ECAPA-TDNN from SpeechBrain~\cite{speechbrain}. These embeddings were then projected onto a 2D space using t-SNE for visualization~\cite{tSNE}. As shown in Figure 4, the plot reveals distinct clusters for individual speakers that remain well-grouped across different compression levels. We observe that intra-speaker variance increases progressively as the bitrate decreases from baseline (blue) to 1.5 kbps (red), indicating that embeddings become less consistent. Moreover, the embeddings that lie furthest from their respective cluster centroids are mostly from the lowest bitrate condition (red), highlighting how extreme compression destroys embedding cohesion. The inter-speaker separation also decreases monotonically with bitrate, suggesting a loss of discriminative features and speaker separability.

\begin{figure}[t]
  \centering
  \includesvg[width=\linewidth]{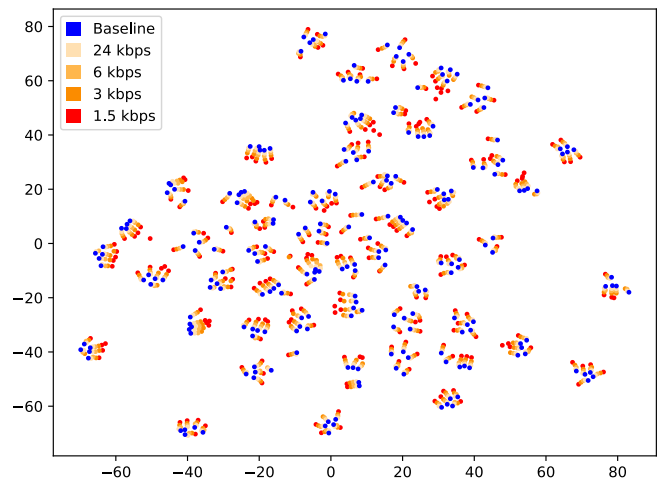}
  \caption{t-SNE of ECAPA-TDNN embeddings for 50 speakers (5 utterances each) across varying bitrates for Encodec. Each cluster represents one speaker, with points colored based on their bitrate (blue $\rightarrow$ red)}
  \label{fig:tsne_visualization}
\end{figure}

\section{Discussion}

\subsection{Comparison with baseline}
All SV models perform best under baseline conditions, achieving the lowest EER/MinDCF. At 1.5 kbps, the EER increases roughly ten-fold relative to the baseline. This dramatic performance drop indicates that critical speaker discriminative features, such as spectral details (formants, harmonics), prosody, phase, and temporal information, are lost at such aggressive compression levels. Performance degrades evenly across all models, regardless of size, because larger architectures cannot leverage their extra capacity when the input lacks most of the fine-grained details. 

In the mid-range (3-12 kbps), we observe a gradual recovery in performance and as the bitrate increases to 24 kbps, the gap with the baseline narrows down, demonstrating the potential for NACs to preserve speaker identity while achieving substantial compression.

\subsection{Comparison across different VoxCeleb1 test splits}
The three VoxCeleb1 test splits show progressively higher difficulty levels, consistent across all SV models and compression levels. The VoxCeleb1-O split represents the easiest scenario, with baseline EER for all models below 1\% (Table III). The extended split, VoxCeleb1-E introduces additional speakers and recording conditions, yielding slightly higher baseline errors.

VoxCeleb1-H is the most difficult split with challenging acoustic conditions and speaker pairs pushing baseline EERs to 1.5–1.9\% (Table V). Under aggressive compression, the relative performance drop is smallest on VoxCeleb1-O and largest on VoxCeleb1-H, indicating that harder sets are more vulnerable to information loss.

\subsection{Comparison across neural audio codecs}
Model performance is strongly influenced by compression level, with the EER increasing significantly as the bitrate decreases, as shown in Figure 3. One contributing factor is that RVQ-based encoding treats all speech features equally, leading to redundant compression of sparse information and limiting low-bitrate performance~\cite{optimizingneuralspeechcodec}.

DAC evaluated at 8 kbps provides a valuable reference point for medium-bitrate performance, delivering similar or even slightly better results compared to Encodec at 24 kbps. This could be attributed to DAC's improved RVQ-GAN~\cite{vq_gan,improved_VQGAN} model which improves codebook usage and bitrate efficiency. This could have helped preserve more speaker-relevant acoustic features even at lower bitrates. It also uses Snake activations~\cite{snake_activations}, which might be better suited for modeling high-frequency and periodic structures like pitch and timbre, helping to preserve speaker identity. DAC also causes the least distortion to genuine samples compared to other NACs~\cite{neuralcodecbasedadversarialsample}. This suggests that codec architecture and training method can be more important than the bitrate alone~\cite{codec-superb}, as DAC achieves a better balance between performance and bandwidth compared to Encodec.

\subsection{Comparison between neural audio codecs and Opus}
When comparing NACs with Opus at different bitrates, we observe an interesting performance divergence at around 12 kbps as shown in Figure 1, revealing their individual strengths and weaknesses in high and low bandwidth regimes. Although NACs show some degradation at higher bitrates, they demonstrate larger advantages in low-bitrate settings.

At higher bitrates, NACs show slightly worse performance compared to Opus. This reflects a design trade-off where NACs are optimized for perceptual quality rather than preserving fine-grained speaker-specific features~\cite{speakeranonymizationusingneural, neuralcodecbasedadversarialsample}. The extra bits at higher quantization levels are allocated towards improving perceptual nuances rather than preserving micro-variances that help an SV model.

Additionally, RVQ codebooks can exhibit ``centroid crowding'', where perceptually similar audio frames from different speakers may be assigned to the same centroid, resulting in a form of compression-induced embedding collapse~\cite{representationcollapsingproblemsvector}. This effect is further compounded by the low-pass frequency response characteristics of RVQ-based audio tokenizers~\cite{discreteaudiorepresentationalternative}, which inherently filter out high-frequency details important for SV tasks. Opus, on the other hand, preserves speaker discriminative features via explicit linear predictive coding (LPC~\cite{itutg7222}) and modified discrete cosine transform (MDCT~\cite{iso13818-7}).

At low-bitrate settings, Opus switches to a narrowband SILK mode, which (unlike wideband mode) discards all energy content above 4 kHz. Most of the fine-grained speaker characteristics such as formant details and high-frequency phase cues are present in this upper-band energy, the loss of which is responsible for its performance degradation. NACs, on the other hand, maintain full-bandwidth representations that preserve relevant speaker cues more effectively, explaining their superior performance at lower bitrates.

\section{Conclusion}
With the current state-of-the-art models, SV performance on transmitted audio remains a balance between bandwidth and performance. Even though NACs do not surpass traditional codecs across the entire bandwidth spectrum, they show huge advantages at lower bitrates, making them an ideal choice for highly bandwidth-constrained scenarios. To bridge the remaining gap at higher bitrates, future work must be focused on improving NAC design and SV model adaptations to explicitly account for speaker identity. In the current scenario, an empirical evaluation needs to be made for any proposed system consisting of SV models and compression methods to determine the best compression and transmission approach.


\printbibliography

\end{document}